\documentclass[english]{article}
\usepackage[latin1]{inputenc}
\usepackage{amsmath}
\usepackage{graphicx}

\makeatletter

\newcommand{\noun}[1]{\textsc{#1}}

\newcommand{\lyxaddress}[1]{
\par {\raggedright #1
\vspace{1.4em}
\noindent\par}
}

\makeatletter

\makeatletter



\makeatother

\makeatother

\usepackage{babel}
\makeatother
\begin{document}

\title{Boosted Kerr black hole}

\author{Janusz Karkowski }

\maketitle

\lyxaddress{Institute of Physics, Jagellonian University, Reymonta 4, 30-059
Krak\'{o}w, Poland}

\begin{abstract}
Initial data for boosted Kerr black hole are constructed in an axially
symmetric case. Momentum and hamiltonian constraints are solved numerically
using finite element method (FEM) algorithms. Both Bowen-York and
puncture boundary conditions are adopted and appropriate results are
compared. Past and future apparent horizons are also found numerically
and the Penrose inequality is tested in detail. 
\end{abstract}

\section{Introduction}

We will construct initial data for single boosted Kerr black hole
in the axially symmetric case with the total momentum parallel to
the angular momentum of the black hole. We assume also asymptotical
flatness. Two popular approaches to the constructions of black hole
initial data will be adopted. The first is the Bowen-York solution
\cite{Bowen} based on the conformal transverse-traceless decomposition
\cite{Cook} of Einstein equations with certain additional assumptions
such as conformal flatness and maximal slicing. The second is the
puncture approach \cite{Brandt} developed by Brandt and Brügmann.
Their idea was to remove analytically the coordinate singularity at
the location of the black hole. In this way they avoided inner boundary
conditions present in the Bowen-York approach. We follow these ideas
as close as possible but some important modifications are necessary.
Garat and Price \cite{Price} have proved nonexistence of conformally
flat slices of the Kerr spacetime. Thus the assumption of conformal
flatness must be abandoned. We assume maximal slicing in order to
simplify the problem but in general it is not necessary. Our approach
is straightforward. We use the quasi-isotropic radial coordinate \cite{Cook},
\cite{Dain} in which it is easy to show that the Kerr solution is
composed of two isometric regions smoothly joined at a sphere of some
radius. Then the original Bowen-York inner boundary conditions can
be applied. The momentum constraints in the axially symmetric case
can be solved analytically for the flat background metric \cite{Frauendiener},
\cite{Malec}, and they can be solved numerically in the case of metrics
which are not conformally flat. We solve the hamiltonian constraint
using the finite element method (FEM) \cite{fem} techniques, then
we find the apparent horizon and test the Penrose inequality.

Our paper consists of three parts. In the first we shortly review
Einstein constraint equations. The second section discusses the numerical
approach and special assumptions that are made there. The third part
deals with detailed results.

\section{Einstein Constraint Equations}

\subsection{ADM framework}

In the Cauchy formulation of Einstein equations \cite{Misner},\cite{Cook}
the whole 4-dimensional manifold is foliated into a set of the so-called
slices. These slices are the 3-dimensional, spacelike surfaces labeled
by a parameter $t$ (time). The Einstein equations are projected onto
these hypersurfaces. One obtains six evolution equations and four
constraint equations that the metric, $\gamma_{ij}$, and the extrinsic
curvature, $K_{ij}$, of each slice must satisfy. In the ADM (3+1)
decomposition of the spacetime the metric can be written as \begin{equation}
ds^{2}=-N^{2}dt^{2}+\gamma_{ij}(dx^{i}+\beta^{i}dt)(dx^{j}+\beta^{j}dt).\label{metric31}\end{equation}
 The proper time between slices is given by the lapse function $N$
and the coordinate drift between slices is described by the shift
vector $\beta^{i}$. The evolution equations in a vacuum have the
following form \begin{eqnarray}
(\partial_{t}-L_{\beta})\gamma_{ij} & = & -2NK_{ij},\label{gammat}\\
(\partial_{t}-L_{\beta})K_{ij} & = & (-\nabla_{i}\nabla_{j}+R_{ij}+KK_{ij}-2K_{ik}K_{j}^{k})N.\label{curvt}\end{eqnarray}
 Here $R_{ij}$ is the Ricci tensor on the slice, $K$ is the trace
of the extrinsic curvature and the covariant derivative $\nabla_{i}$
is taken with respect to the spatial metric $\gamma_{ij}$. $L_{\beta}$
denotes the Lie derivative along the shift vector $\beta^{i}$.

The minimal set of initial data consisting of the 3-dimensional metric
$\gamma_{ij}$ and the extrinsic curvature $K_{ij}$ cannot be specified
independently on the initial slice. These data are constrained by
four equations, as mentioned above, because the slices must fit properly
into the 4-dimensional manifold. The first of these equations is known
as the hamiltonian (or scalar) constraint \begin{equation}
R+K^{2}-K_{ij}K^{ij}=0,\label{hamil_con}\end{equation}
 and the other three, called momentum (or vector) constraints read
as \begin{equation}
\nabla_{j}(K^{ij}-\gamma^{ij}K_{l}^{l})=0.\label{mom_con}\end{equation}
 Here $R$ denotes the Ricci scalar. Let us note that if the above
constraint equations are satisfied on one initial hypersurface, the
same holds true on each slice. It is also important that the constraint
equations do not depend on the lapse function $N$ and the shift vector
$\beta^{i}$ which can be freely chosen on each slice. This is the
base of the (general) gauge invariance of the theory.

\subsection{York-Lichnerowicz conformal approach}

The process of solving of the constraint equations can be significantly
simplified by the conformal York-Lichnerowicz decomposition \cite{Cook},
\cite{Bowen}: \begin{eqnarray}
\gamma_{ij} & = & \psi^{4}\hat{\gamma}_{ij},\label{lich1}\\
K_{ij} & = & \hat{K}_{ij}\psi^{-2}.\label{lich2}\end{eqnarray}
 The new metric $\hat{\gamma}_{ij}$ is called the conformal (or background)
metric and can be chosen arbitrarily. This conformal tranformation
preserves the shape of the momentum constraints: \begin{equation}
\hat{\nabla}_{j}(\hat{K}^{ij}-\hat{\gamma}^{ij}\hat{K}_{l}^{l})=0,\label{york2}\end{equation}
 and the hamiltonian constraint can be rewritten as \begin{equation}
\hat{\nabla}^{2}\psi=\frac{1}{8}\hat{R}\psi-\frac{1}{8}(\hat{K}_{ij}\hat{K}^{ij}-(\hat{K}_{i}^{i})^{2})\psi^{-7}.\label{york1}\end{equation}
 The covariant derivative $\hat{\nabla}$ and the Ricci scalar $\hat{R}$
correspond to the background metric $\hat{\gamma}_{ij}$. We will
also impose the maximal slicing condition \begin{equation}
K=K_{i}^{i}=0.\label{maxsli}\end{equation}
 This equality can be understood as a gauge fixing condition. It fixes
the lapse function $N$ and is convenient in avoiding coordinate singularities
during the time evolution of the system. It also simplifies the constraint
equations which finally take the form

\begin{equation}
\hat{\nabla}_{j}\hat{K}^{ij}=0,\label{mom_con_final}\end{equation}

\begin{equation}
\hat{\nabla}^{2}\psi=\frac{1}{8}\hat{R}\psi-\frac{1}{8}\hat{K_{ij}}\hat{K}^{ij}\psi^{-7}.\label{ham_con_final}\end{equation}

\subsection{Axially symmetric ansatz}

We would like to solve numerically the constraint equations for the
boosted Kerr black hole with momentum directed along angular momentum.
Therefore it is desirable to choose a Weyl metric as the background
metric. The Weyl metric is axially symmetric and its line element
in the spherical coordinates has the form \cite{Brill}, \cite{Koc}
\begin{equation}
\hat{\gamma}_{ij}dx^{i}dx^{j}=exp(-q(r,\theta))(dr^{2}+r^{2}d\theta^{2})+r^{2}\sin^{2}\theta d\phi^{2}.\label{wyelm}\end{equation}
 Let us recall that the Kerr initial data can be written (in terms
of quasi-isotropic radial coordinate $r$) as \cite{Cook} \begin{eqnarray}
exp(-q(r,\theta)) & = & \frac{\Sigma^{2}}{(R^{2}+a^{2})^{2}-\Delta a^{2}\sin^{2}\theta},\label{qwyel}\\
\Sigma & = & R^{2}+a^{2}\cos^{2}\theta,\\
\Delta & = & R^{2}-2mR+a^{2},\\
R & = & r+m+\frac{b^{2}}{r},\\
b & = & \frac{\sqrt{m^{2}-a^{2}}}{2},\label{inver}\end{eqnarray}
 where $m,a$ are Kerr parameters and the conformal factor $\psi(t,\theta)$
is given by \begin{equation}
\psi^{4}(r,\theta)=\frac{\Sigma}{r^{2}}exp(q(r,\theta)).\label{kerrsol}\end{equation}
 We will fix the $exp(-q(r,\theta))$ factor in (\ref{wyelm}) to
be that from the Kerr metric (\ref{qwyel}). The external curvature
in the axially symmetric case with maximal slicing can be written
in the form \begin{equation}
(\hat{K}_{ij})=\left(\begin{array}{ccc}
\frac{f_{11}(r,\theta)}{r^{2}} & \frac{f_{12}(r,\theta)\sin\theta}{r} & \frac{f_{13}(r,\theta)}{r^{2}}\\
\ldots & f_{22}(r,\theta) & \frac{f_{23}(r,\theta))}{\sin\theta}\\
\ldots & \ldots & -\frac{(f_{11}(r,\theta)+f_{22}(r,\theta))\sin^{2}\theta}{exp(-q(r,\theta))}\end{array}\right),\label{curv}\end{equation}
 where $i,j=r,\theta,\phi$ and $\hat{K}_{ij}=\hat{K}_{ji}$. The
momentum constraint equations (\ref{mom_con_final}) reduce then to
\begin{eqnarray}
 & r\partial_{r}(rf_{11})+\frac{1}{2}r^{2}(f_{11}+f_{22})\partial_{r}q-\partial_{y}((1-y^{2})rf_{22})=0,\label{wyelk1}\\
 & \partial_{r}((1-y^{2})rf_{12})-\frac{1-y^{2}}{2}(f_{11}+f_{22})\partial_{y}q-\partial_{y}((1-y^{2})f_{22})+yf_{11}=0,\label{wyelk2}\\
 & \partial_{r}f_{13}+\partial_{\theta}f_{23}=0.\label{wyelk3}\end{eqnarray}
 The axial symmetry allows one to solve these equations by the method
of separating variables. The functions $f_{13}$ and $f_{23}$ are
present in the third equation only. This equation can be easily integrated.
The result is \begin{eqnarray}
f_{13} & = & -\partial_{\theta}Z(r,\theta),\\
f_{23} & = & \partial_{r}Z(r,\theta),\end{eqnarray}
 where $Z(r,\theta)$ is an arbitrary function. If we put \begin{eqnarray}
f_{11} & = & f_{12}=f_{22}=0,\\
Z(r,\theta) & = & ma(\cos^{3}\theta-3\cos\theta)-\frac{ma^{3}\sin^{4}\theta\cos(\theta)}{\Sigma},\end{eqnarray}
 then we recover the full Kerr initial data.

\section{Numerical approach}

\subsection{Momentum constraints}

The background metric becomes flat for $q(r,\theta)=0$. This is the
case when the angular momentum of the Kerr black hole is equal to
zero and we get the Schwarzschild solution. Thus for $a=0$ we have
the flat background metric and then the momentum constraints can be
solved analytically (assuming maximal slicing and axial symmetry).
The solution reads as \cite{Karkowski} \begin{eqnarray}
\hat{K}_{rr} & = & \frac{1}{r^{3}}\partial_{y}^{2}W(r,\theta),\label{dain1}\\
\hat{K}_{r\theta} & = & \frac{1}{r\sin\theta}\partial_{y}\partial_{r}W(r,\theta),\label{dain2}\\
\hat{K}_{\theta\theta} & = & \frac{1}{\sin^{2}\theta}(\partial_{r}(r\partial_{r}W(t,\theta))+\frac{1}{r}(y\partial_{y}W(r,\theta)-W(r,\theta))),\label{dain3}\\
\hat{K}_{r\phi} & = & \frac{1}{r^{2}}\partial_{y}Z(r,\theta),\label{dainz1}\\
\hat{K}_{\theta\phi} & = & \frac{1}{\sin\theta}\partial_{r}Z(r,\theta),\label{dainz2}\end{eqnarray}
 where $y=\cos\theta$ and $W,Z$ are arbitrary functions. The component
$\hat{K}_{\phi\phi}$ can be calculated from the maximal slicing condition
$\hat{K}_{i}^{i}=0$. These formulae can be obtained as a special
case of Dain-Friedrich \cite{Friedrich} conformally flat initial
data that have been written in the Newman-Penrose formalism, or more
simply by a direct integration of the momentum constraint equations
(\ref{wyelk1}-\ref{wyelk3}).

Our idea is to obtain the momentum constraint solutions as a deformation
of the above flat background formulae with $a$ as the deformation
parameter. Thus we look for solutions in the form \begin{equation}
\hat{K}_{ij}(a)=\hat{K}_{ij}(a=0)+\delta\hat{K}_{ij}(a).\end{equation}
 where $\hat{K}_{ij}(a=0)$ are given above and $\delta\hat{K}_{ij}(a)$
will be found numerically from (\ref{curv}),(\ref{wyelk1}-\ref{wyelk2})
with appropriate boundary conditions which will be discussed below.

\subsection{Hamiltonian constraint}

Numerical methods are necessary in order to solve the hamiltonian
constraint in both flat and Weyl background metric. This constraint
is a quasilinear elliptic equation and can be solved by the standard
Newton method. The equation has the form \begin{eqnarray}
L\psi & = & F(r,\theta,\psi),\label{newtong1}\\
F(r,\theta,\psi) & = & f(r,\theta)\psi^{-7},\label{newtong2}\end{eqnarray}
 where L is a linear operator and $F,f$ are arbitrary functions.
Let $\psi_{n}$ be the n-th approximation in the Newton sequence.
Then the following reccurence formula can be easily obtained \begin{eqnarray*}
 & (L-F_{\psi}(r,\theta,\psi_{n}))\delta\psi_{n}=F(r,\theta,\psi_{n})-F(r,\theta,\psi_{n-1}),\\
 & -F_{\psi}(r,\theta,\psi_{n-1})(\psi_{n}-\psi_{n-1}).\end{eqnarray*}
 Here $\delta\psi_{n}$ is a correction to the n-th approximation
and $F_{\psi}$ denotes the partial derivative of $F$ with respect
to $\psi$. It is important that the conformal factor $\psi$ (\ref{kerrsol})
is the analytical solution of (\ref{york1}) for Kerr initial data.
Therefore it is an excellent example for testing of numerical methods
and can serve as the 0-th order approximation to more complicated
cases.

\subsection{Bowen-York boundary conditions}

Bowen and York demand the metric $\gamma_{ij}$ and the external curvature
$K_{ij}$ to be invariant (up to a sign) under inversion through a
sphere of radius $b$ \cite{Bowen}, \cite{Dain}. This transformation
is given in spherical coordinates by \begin{equation}
\bar{r}=\frac{b^{2}}{r},\bar{\theta}=\theta,\bar{\phi}=\phi.\label{invtrans}\end{equation}
 The metric $\gamma_{ij}$ is invariant if the background metric $\hat{\gamma}_{ij}$
is flat and if the conformal factor $\psi(r,\theta,\phi)$ satisfies
\begin{equation}
\psi(r,\theta,\phi)=\frac{b}{r}\psi(\bar{r},\bar{\theta},\bar{\phi}).\label{psis}\end{equation}
 The same holds true for Weyl background metric (\ref{wyelm}) if
the function $q(r,\theta)$ is invariant under the inversion transformation.
This is the case for the Kerr metric with $b$ given in (\ref{inver}).
The invariance of the extrinsic curvature is equivalent to the following
transformation rules \begin{eqnarray}
\text{} & \hat{K}_{rr}(r,\theta,\phi)=\pm\frac{b^{6}}{r^{6}}\hat{K}_{rr}(\bar{r},\bar{\theta},\bar{\phi}),\nonumber \\
 & \hat{K}_{r\theta}(r,\theta,\phi)=\mp\frac{b^{4}}{r^{4}}\hat{K}_{r\theta}(\bar{r},\bar{\theta},\bar{\phi}),\nonumber \\
 & \hat{K}_{\theta\theta}(r,\theta,\phi)=\pm\frac{b^{2}}{r^{2}}\hat{K}_{\theta\theta}(\bar{r},\bar{\theta},\bar{\phi}),\label{pawel}\\
 & \hat{K}_{\theta\phi}(r,\theta,\phi)=\mp\frac{b^{2}}{r^{2}}\hat{K}_{\theta\phi}(\bar{r},\bar{\theta},\bar{\phi}),\nonumber \\
 & \hat{K}_{\phi\phi}(r,\theta,\phi)=\pm\frac{b^{2}}{r^{2}}\hat{K}_{\phi\phi}(\bar{r},\bar{\theta},\bar{\phi}).\nonumber \end{eqnarray}
 We construct a family of invariant solutions of momentum constraints
(\ref{mom_con_final}) for the flat background metric by putting the
generating function $W(r,\theta)$ in formulae (\ref{dain1}-\ref{dain3})
equal to \begin{equation}
\partial_{y}W(r,\theta)=-\frac{3}{2}P(r\pm\frac{b^{2}}{r})\sin^{2}\theta g(\theta),\label{wgen1}\end{equation}
 where $P$ is a parameter and $g(\theta)$ an arbitrary function.
Let us note that for $g(\theta)=1$ a Bowen-York solution is usually
written in cartesian coordinates in the form \cite{Bowen}

\begin{equation}
\hat{K}_{ij}=\frac{3}{2r^{2}}(P_{i}n_{j}+P_{j}n_{i}-(\eta_{ij}-n_{i}n_{j})P^{k}n_{k})\mp\frac{3b^{2}}{2r^{4}}(P_{i}n_{j}+P_{j}n_{i}-(\eta_{ij}-5n_{i}n_{j})P^{k}n_{k}),\label{BYcurve}\end{equation}
 where $n_{i}$ is the unit normal to a sphere $r=const$. In this
case $P_{i}$ is the total linear momentum \begin{equation}
P_{i}=\frac{1}{8\pi}\int K_{ij}d^{2}S^{j},\label{fullmomentum}\end{equation}
 directed along z-axis. The appropriate formulae for the Weyl background
metric result from (\ref{wgen1}) plus some corrections (which become
zero for $a=0$) that satisfy symmetry conditions (\ref{pawel}).
We get the following boundary conditions for corrections to the extrinsic
curvature functions (\ref{wyelk2}) \begin{eqnarray*}
 & \partial_{r}\delta f_{11}(b,\theta)+\frac{1}{b}\delta f_{11}(b,\theta)=0,\\
 & \delta f_{12}(b,\theta)=0,\\
 & \lim_{r\rightarrow\infty}\delta f_{11}(r,\theta)=\lim_{r\rightarrow\infty}\delta f_{12}(r,\theta)=0.\end{eqnarray*}
 The function $f_{22}$ can be chosen arbitrary. It is reasonable
to keep the same relation between $f_{11}$ and $f_{22}$ as in the
flat background case, that is \[
f_{11}+\frac{1}{2}f_{22}=0.\]
 The boudary conditions for the conformal factor $\psi$ result from
(\ref{psis}) (and asymptotical flatness) and read as \begin{eqnarray*}
 & \partial_{r}\psi(b,\theta)+\frac{1}{2b}\psi(b,\theta)=0,\\
 & \lim_{r\rightarrow\infty}\psi(r,\theta)=1.\end{eqnarray*}

\subsection{Puncture boundary conditions}

In the Bowen-York approach each slice consists of two isomorphic ends
(the inversion transformation (\ref{invtrans}) defines an isometry
of the physical metric) separated by a sphere of radius $b$ and we
can limit ourselves to one end without any singularities. The puncture
method takes care of singularities in another way. In the flat background
case we start with the curvature generated by \begin{equation}
\partial_{y}W(r,\theta)=-\frac{3}{2}Pr\textrm{sin}^{2}\theta g(\theta),\label{wgenp}\end{equation}
 and continue with the conformal factor of the form

\begin{equation}
\psi=1+\frac{m}{2r}+\delta\psi,\label{psipunc}\end{equation}
 where $\delta\psi$ is assumed to be nonsingular on the whole initial
slice. Inserting (\ref{psipunc}) into (\ref{york1}) it can be easily
seen that the admissible singularities (at $r=0$) in curvature functions
are of type at most $r^{-3}$. That is why the term of order $r^{-1}$
from (\ref{wgen1}) had to be excluded in (\ref{wgenp}).

The above formulae can be generalized to the Kerr background case
in the following manner. The flat curvature functions are completed
by ($a$-dependent) corrections in order to satisfy momentum constraint
equations (\ref{wyelk2}). These corrections are expected to be nonsingular.
The form of the conformal factor (\ref{psipunc}) is changed to \begin{equation}
\psi_{punc}=\psi+\delta\psi,\label{psipunct}\end{equation}
 where $\psi$ is given in (\ref{kerrsol}). The correction $\delta\psi$
must also be nonsingular. This becomes clear if we compare (\ref{psipunc})
with the asymptotic expansions of $\psi$ near $r=0$ and $r=\infty$
\begin{eqnarray*}
 & \psi(r\approx0,\theta)=\frac{\sqrt{m^{2}-a^{2}}}{2r}+\frac{m}{\sqrt{m^{2}-a^{2}}}+\frac{a^{2}r}{\sqrt{m^{2}-a^{2}}^{3}}+O(r^{2}),\\
 & \psi(r\rightarrow\infty,\theta)=1+\frac{m}{2r}+\frac{1}{8}\frac{a^{2}}{r^{2}}+O(\frac{1}{r^{3}}).\end{eqnarray*}
 Let us note that the boundary conditions for $r\rightarrow\infty$
are the same as in the Bowen-York case, for instance \[
\lim_{r\rightarrow\infty}\delta\psi(r,\theta)=0.\]

\section{Numerical results}

\subsection{Numerical techniques}

The first step is to find the mapping of the radial coordinate onto
some compact interval. In the Bowen-York case we have used a new coordinate
$x$ defined by

\[
x=\frac{b}{r},\]
 which mapps the interval $[b,\infty)$ onto $[0,1]$. Similarly,
for puncture boundary conditions we have put

\[
x=\frac{r}{1+r}.\]
 In both cases we have to solve elliptic partial differential equations
on the rectangle $[[0,1],[-1,1]]$ with mixed boundary conditions.
These equations for extrinsic curvature and conformal factor have
been solved on a dense lattice with 5000x200 points with the help
of excellent numerical algorithms for sparse linear systems: MUMPS
\cite{mumps}, UMFPACK \cite{umfpack} and HYPRE \cite{hypre}. The
precision was tested on the conformal factor of the Kerr solution
(\ref{kerrsol}), and for instance the mass parameter was recovered
with the relative error less than $10^{-4}$.

We have achieved still better precision using finite elements method
(FEM). This method is based on triangulations which can be fitted
to a shape of succesive approximations of the solution. Especially
powerfull has turned out to be the program FreeFem++ \cite{fem} which
is an implementation of the special language dedicated to the finite
element method. We have used it extensively as it enables one to solve
easily PDE problems, both elliptic and time dependent. The relative
numerical precision has been improved up to $10^{-6}$.

\subsection{Detailed results}

Our boundary conditions assume the slices to be asymptotically flat.
The ADM mass is defined as

\begin{equation}
m_{ADM}=\sqrt{E^{2}-P^{i}P_{i}},\label{adm}\end{equation}
 where E is the total energy of the system

\begin{equation}
E=-\frac{1}{2\pi}\int_{S_{\infty}}d^{2}S^{i}\nabla_{i}\psi=\lim_{r\rightarrow\infty}2r(\psi(r,\theta)-1),\label{energia}\end{equation}
 and $P_{i}$ the total momentum defined in (\ref{fullmomentum}).
The ADM mass is a parameter which shows how quickly a slice becomes
flat at infinity. The minimal surface is in turn a closed surface
which locally has a minimal area (in the riemannian geometry). The
following inequality has been proved for slices with the vanishing
extrinsic curvature \cite{Huisken}, \cite{Bray}

\begin{equation}
m_{ADM}\ge\sqrt{\frac{S}{16\pi}}.\label{penrose}\end{equation}
 Here $m$ is the ADM mass and $S$ the area of the outermost minimal
surface. This is the so-called Riemannian version of the Penrose inequality
\cite{Penrose}, \cite{Gibbons}. The minimal surface can be found
from the condition

\begin{equation}
\nabla_{i}n^{i}=0,\end{equation}
 where $n^{i}$ is the unit vector normal to the surface. The Penrose
inequality must be reformulated for slices with non-vanishing extrinsic
curvature. The minimal surface is then replaced by an apparent horizon
by which is understood a closed two-dimensional surface which obeys
one of the two equations \cite{Malec}

\begin{equation}
\theta_{\pm}=\nabla_{i}n^{i}\mp K_{ij}n^{i}n^{j}=0,\end{equation}
 where $\pm$ correspond to the past and future apparent horizons
respectively, and $\theta_{\pm}$ are called optical scalars. In the
axially symmetric case these apparent horizons satisfy the differential
equations \cite{Bowen}

\begin{eqnarray}
 & r_{\theta\theta}+\frac{r_{\theta}^{3}}{r^{2}}(\frac{4\psi_{\theta}}{\psi}-\frac{q_{\theta}}{2}+\cot\theta)-r_{\theta}^{2}(\frac{4\psi_{r}}{\psi}-\frac{q_{r}}{2}+\frac{3}{r})\nonumber \\
 & +r_{\theta}(\frac{4\psi_{\theta}}{\psi}-\frac{q_{\theta}}{2}+\cot\theta)-r^{2}(\frac{4\psi_{r}}{\psi}-\frac{q_{r}}{2}+\frac{2}{r})=\\
 & \pm\frac{1}{\psi^{4}}\sqrt{1+\frac{r_{\theta}^{2}}{r^{2}}}(r^{2}\hat{K}_{rr}+\frac{r_{\theta}^{2}}{r^{2}}\hat{K}_{\theta\theta}-2r_{\theta}\hat{K}_{r\theta}),\nonumber \end{eqnarray}
 with the following boundary conditions for a function $r(\theta)$
at $\theta=0$ and $\theta=\pi$

\begin{equation}
r_{\theta}\vert_{\theta=0}=r_{\theta}\vert_{\theta=\pi}=0.\end{equation}
 The Penrose conjecture for a slice with a non-vanishing extrinsic
curvature has the same form (\ref{penrose}) as before, but $S$ is
the area of the outermost apparent horizon to the future instead of
the minimal surface. The above formulation is one of a number of possible
wordings of the hypothesis - see \cite{Malec} for an extensive discussion
of various formulations of the Penrose conjecture. In fact all the
versions described in \cite{Malec} have been checked. Since the results
differ very little and the hypotheses always stand the tests, we report
here only data concerning the Penrose inequality described above.
This inequality is not proved as yet although there are some partial
results \cite{Swiercz}, \cite{Omur}, \cite{Iriondo}, \cite{Hayward},
\cite{Jezierski}, \cite{Frauendiener} and schemes for the proof
\cite{MMS}, \cite{Roszko}. It is widely believed that if the Penrose
inequality is not true then the cosmic censorship will be broken.

Our numerical program was the following one

\begin{figure}[p]
\includegraphics{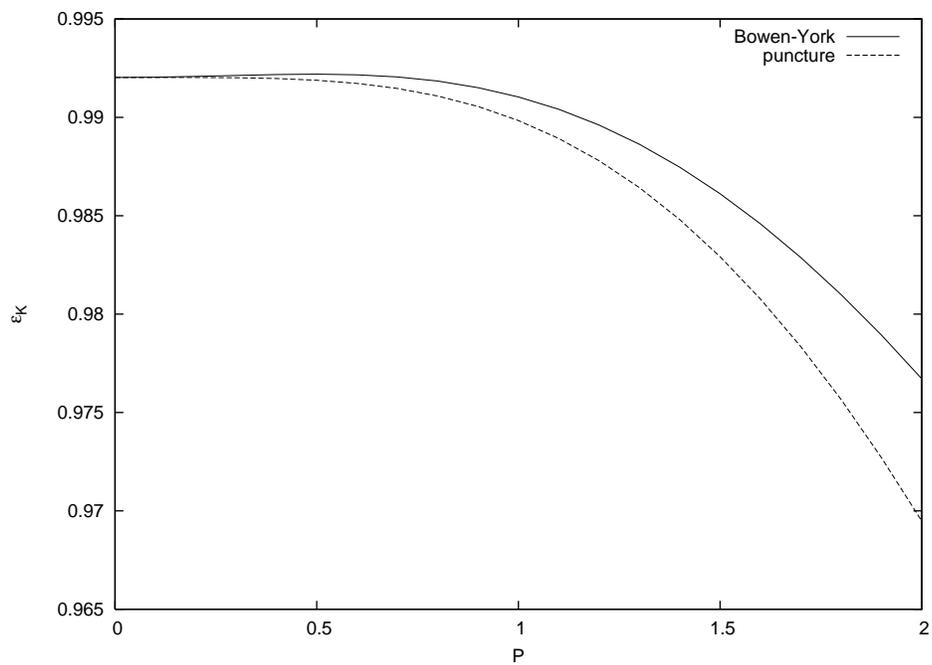}

\caption{The momentum dependence of $\epsilon_{K}$ for $a=0.5$ and $m=2.$}

\label{K05} 
\end{figure}

\begin{figure}
\includegraphics{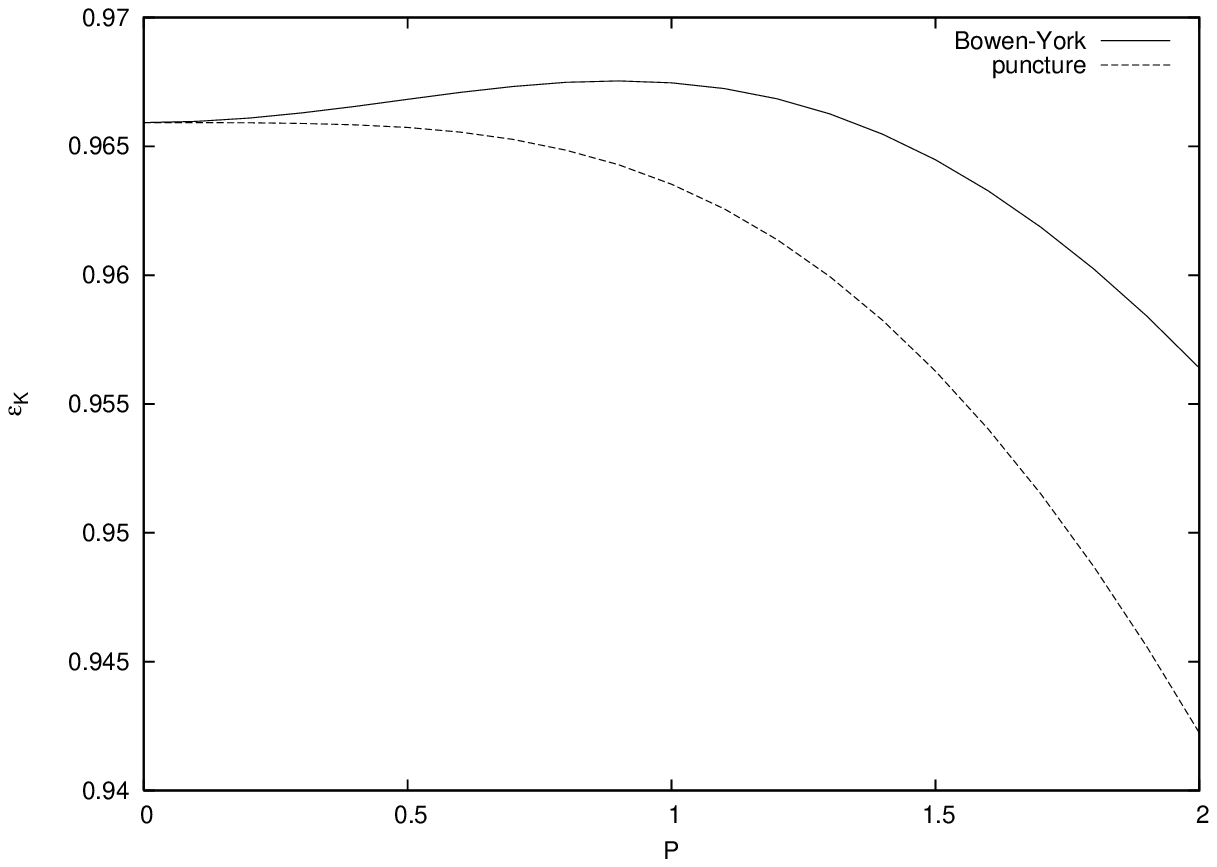}

\caption{The momentum dependence of $\epsilon_{K}$ for $a=1.0$ and $m=2.$}

\label{K10} 
\end{figure}

\begin{figure}
\includegraphics{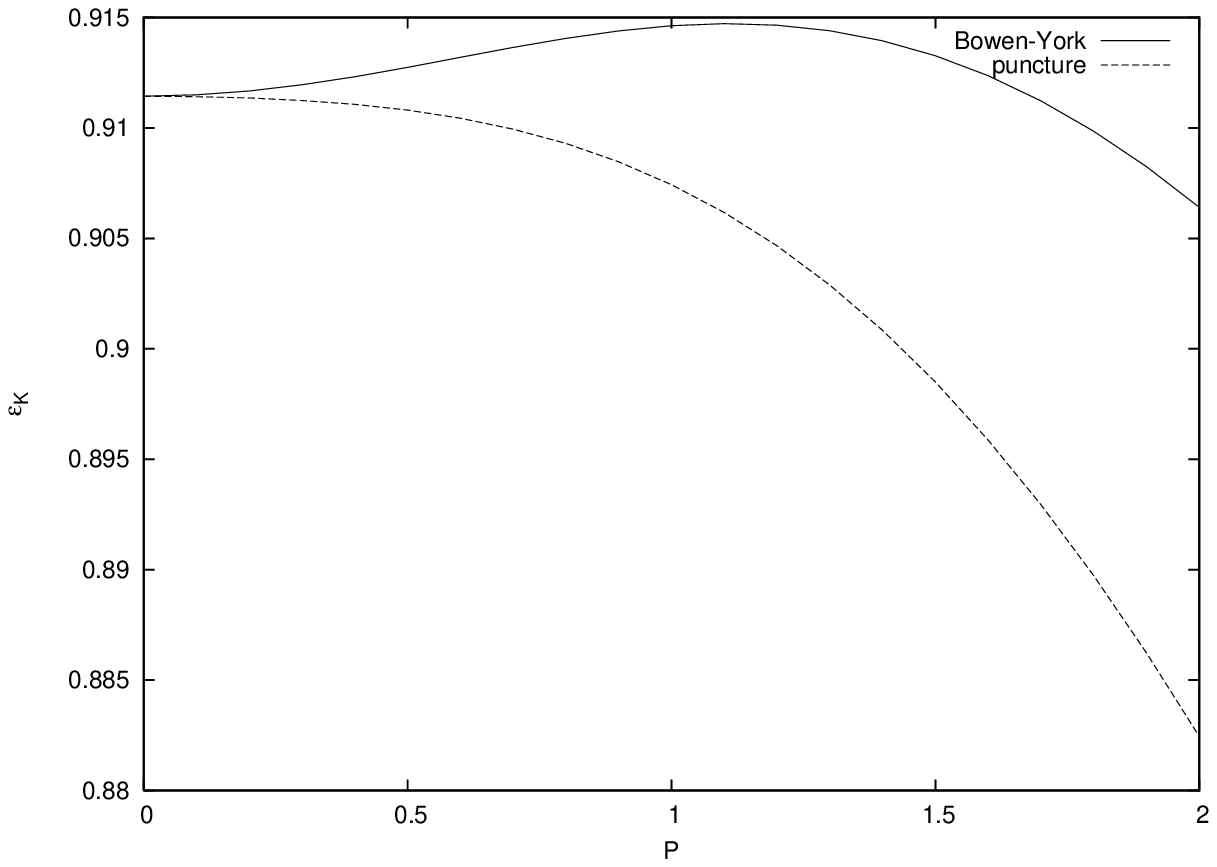}

\caption{The momentum dependence of $\epsilon_{K}$ for $a=1.5$ and $m=2.$}

\label{K15} 
\end{figure}

\begin{itemize}
\item for given values of Kerr parameters $m,a$ solve the momentum constraints, 
\item calculate the conformal factor $\psi$ for several values of the black
hole momentum $P=\sqrt{P_{i}P^{i}}$ (\ref{fullmomentum}), 
\item for each $P$ find the ADM mass, an apparent horizon and its area
\[
S=2\pi\int_{0}^{a}\psi^{4}\exp(-\frac{q}{2})\sqrt{r_{\theta}^{2}+r^{2}}r\sin\theta d\theta,\]

\item graphically show the momentum dependence of the coefficient\[
\epsilon_{K}=\sqrt{\frac{S}{16\pi m_{ADM}^{2}}},\]
 which according to the Penrose inequality should satisfy the inequality
$\epsilon_{K}\leq1$. 
\end{itemize}
The numerical results are presented on three figures (Fig. \ref{K05}-\ref{K15})
. Each of them shows how $\epsilon_{K}$ depends on momentum in case
of Bowen-York and puncture boundary conditions.

Let us note that this dependence is quite different for the Bowen-York
and the puncture boundary conditions. In the Bowen-York case the coefficient
$\epsilon_{K}$ reaches some maximum, that is evident especially for
big values of $a$.

\begin{figure}
\includegraphics{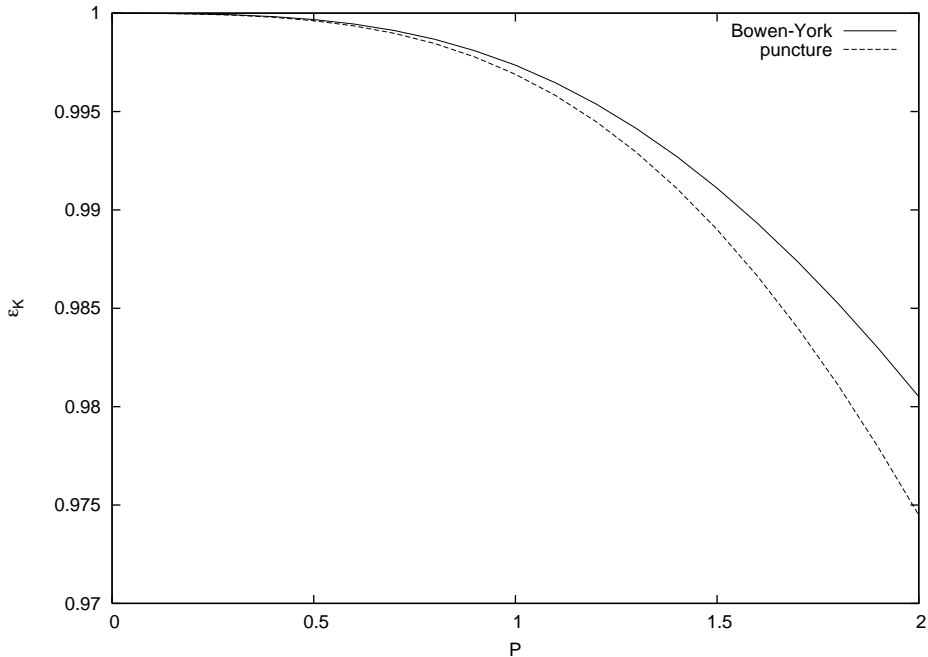}

\caption{The momentum dependence of $\epsilon_{A}$ for $a=0.5$ and $m=2.$}

\label{A05} 
\end{figure}

\begin{figure}
\includegraphics{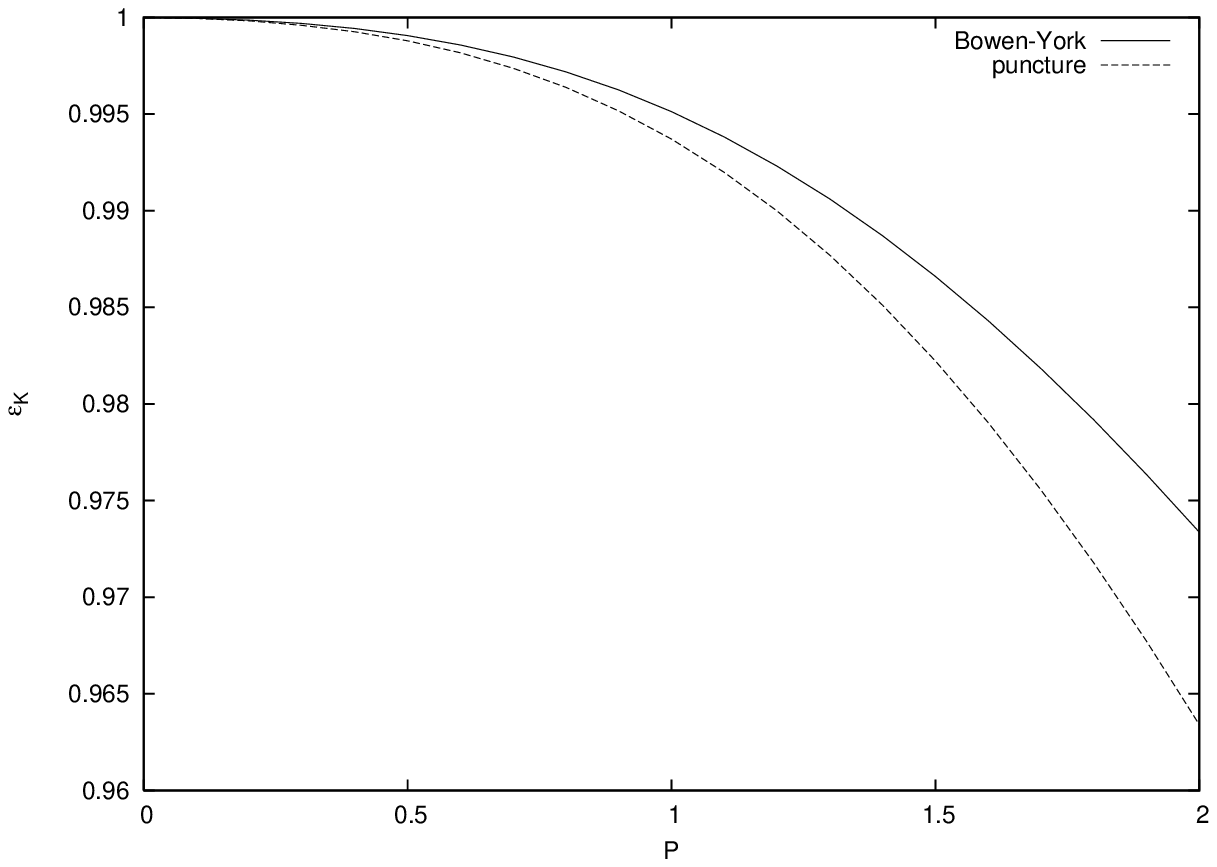}

\caption{The momentum dependence of $\epsilon_{A}$ for $a=1.0$ and $m=2.$}

\label{A10} 
\end{figure}

\begin{figure}
\includegraphics{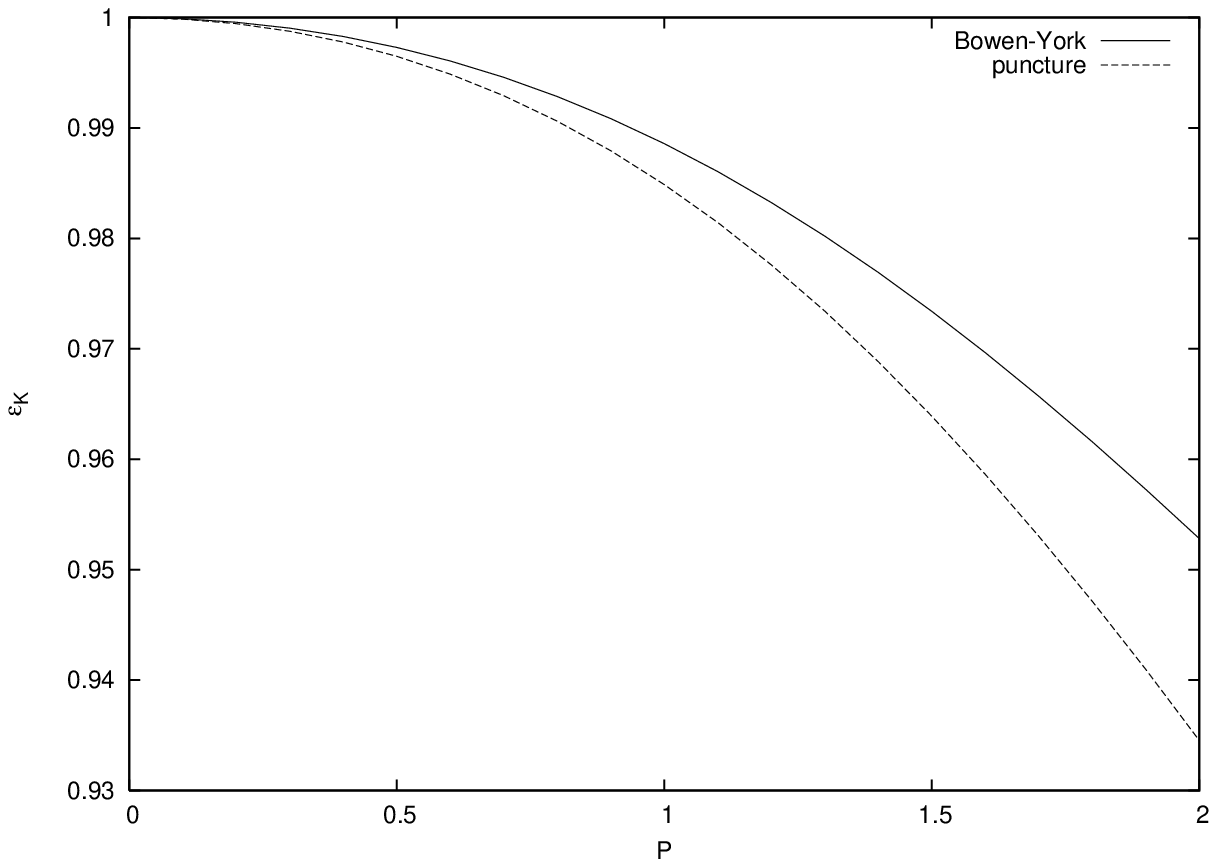}

\caption{The momentum dependence of $\epsilon_{A}$ for $a=1.5$ and $m=2.$}

\label{A15} 
\end{figure}

There is also another inequality which is expected to hold for every
axially symmetric, asymptotically flat initial data. Similarly to
the Penrose inequality we can define a coefficient $\epsilon_{A}$

\begin{equation}
\epsilon_{A}=\sqrt{\frac{S}{8\pi(m_{ADM}^{2}+\sqrt{m_{ADM}^{4}-J^{2}})}},\label{axpenrose}\end{equation}
 where $J$ is the total angular momentum. It is important to note
that in our case

\begin{equation}
J=ma,\label{angular}\end{equation}
 even if the total momentum is not zero. The axially symmetric counterpart
of the Penrose inequality can be stated as

\[
\epsilon_{A}\leq1,\]
 and the equality is reached only for Kerr slices \cite{Hawking}.
The momentum dependence of the coefficient $\epsilon_{A}$ both for
Bowen-York and puncture data is shown on Fig. \ref{A05}-\ref{A15}
.

Let us stress that our results depend only of two pameters $a=\frac{J}{m}$
and $\frac{P}{m}$. Therefore we have limited ourselves to $m=2$.

\section{Conclusions}

We have generalized Bowen-York and puncture constructions of black
hole initial data to the case of a boosted Kerr black hole in the
axially symmetric case. These initial data smoothly depend on the
black hole angular momentum and for $a=0$ coincide with the conformally
flat solutions. Our initial data are very precise and can serve as
the starting point in constructing initial data for binary black holes.
It would be also interesting to compute the long-term numerical evolution
of the single boosted Kerr black hole and look for the quasinormal
modes present in the radiation. We have calculated numerically the
apparent horizons and tested the famous Penrose inequality together
with its more restrictive version valid for axially symmetric initial
data only. The numerical results are in agreement with theoretical
predictions and clearly confirm the fact that the equality in the
axially symmetric version is possible only for non-boosted Kerr black
hole.

Our numerical calculations were mostly based on finite element method
(FEM) techniques. In our opinion these methods lead to more precise
results than those made on rectangular grids. The strength of FEM
lies in the fact that the FEM algorithms have been widely tested in
other areas of physics especially in elasticity, aerodynamics, electrostatics
and in time dependent problems in hydrodynamics.

We believe that our results can be generalized in a straightforward
manner to the case of binary black hole initial data. The work concerning
appropriate puncture construction is in progress.

\paragraph{Acknoledgement. }

The author thanks prof. E. Malec for careful reading of the manuscript
and helpfull comments and suggestions. This work was partly supported
by the Polish State Committee for Scientific Research (KBN) grant
1 PO3B 01229.

\end{document}